\documentclass[aps,prd,onecolumn,showkeys,showpacs,amssymb]{revtex4}
\usepackage{float}
\usepackage{amssymb}
\usepackage{graphicx}
\usepackage{amsmath}
\usepackage{hyperref}

\newcommand{\be}{\begin{equation}}
\newcommand{\ee}{\end{equation}}

\begin{document}

\title{The possible resolution of Boltzmann brains problem in phantom cosmology}

\author{Artyom V. Astashenok, Artyom V. Yurov, Valerian V. Yurov}
\affiliation{I. Kant Baltic Federal University,
\\Institute of
Physics and Technology, 236041, 14, Nevsky st., Kaliningrad,
Russia}

\begin{abstract}
We consider the well-known Boltzmann brains problem in frames of
simple phantom energy models with little rip and big rip
singularity. It is showed that these models (i) satisfy to
observational data and (ii) may be free from Boltzmann brains
problem. The human observers in phantom models can exist only in
during for a certain period $t<t_{f}$ ($t_{f}$ is lifetime of
universe) via Bekenstein bound. If fraction of unordered observers
in this part of universe history is negligible in comparison with
ordered observers than Boltzmann brains problem doesn't appear.
The bounds on model parameters derived from such requirement don't
contradict to allowable range from observational data.
\end{abstract}



\maketitle

\section{Introduction}

The discovery of accelerated expansion of the universe \cite{Riess},
\cite{Perlmutter} led to number of new ideas/solutions in cosmology. For explanation of the cosmic acceleration the various models of so-called dark energy are proposed (for recent reviews, see \cite{Dark-1}, \cite{Dark-2}, \cite{Dark-3}, \cite{Dark-4}, \cite{Dark-5}, \cite{Dark-6}). For dark energy the equation-of-state parameter is negative:
\be
w=p/\rho<0\, ,
\ee
where $\rho$ is the dark energy density and $p$ is the pressure.

In principle recent observations favor to the standard
cosmological model ($\Lambda$CDM-model) with a universe made up
71.3\% of vacuum energy ($w=-1$) and only 27.4\% of a combination
of dark matter and baryonic matter \cite{Kowalski}. In
observational astrophysics the simple dark energy model
($w=w_{0}$) is usually considered as alternative to
$\Lambda$CDM-model. In frames of this model the latest
cosmological data give for $w=-1.04^{+0.09}_{-0.10}$
\cite{PDP,Amman}.

If $w<-1$ the violation of all four energy conditions  occurs. The
corresponding phantom field, which is instable as quantum field
theory \cite{Carrol} but could be stable in classical cosmology
may be naturally described by the scalar field with the negative
kinetic term.

The model with constant $w<-1$ leads to Big Rip singularity \cite{Starobinsky}, \cite{Caldwell}, \cite{Frampton}, \cite{Diaz}, \cite{Nojiri}. One note that condition $w<-1$ is not sufficient for a
singularity occurrence. Moreover, one can construct such models in which $w$ asymptotically
tends to $-1$ and energy density increases with time or remains constant but
there is no finite-time future singularity \cite{Nojiri-3}, \cite{Stefanic}, \cite{Frampton-2}. Of course, most evident case is when Hubble rate
tends to  constant  (cosmological constant or asymptotically de Sitter space).
Very interesting situation is related with Little Rip cosmology
\cite{Frampton-2} where Hubble rate tends to infinity in the infinite future
(for further investigation, see \cite{Frampton-3}, \cite{Wei}, \cite{Chimento}, \cite{Cai}, \cite{ito}, \cite{Frampton-4}).
The key point is that if $w$ approaches $-1$ sufficiently rapidly,
then it is possible to have a model in which the time required for singularity
is infinite, i.e., singularity effectively does not occur. Nevertheless, it
may be shown  that even in this case the disintegration of bound structures
takes place in the way similar to Big Rip.

There are many advantages for this model and we shall consider
one of them: such models may be free from a problem of
Boltzmann brains (BB).

In \cite{page-1} Page has shown that if the dS universe will
expand no less then $10^{60}$ yr (with the present value of the
Hubble root $H_0=72\pm 8$ km/s/Mpc) then such universe will be
filled with BB: the predominant race in the
universe is  BB rather then ordered observers. On the other hand,
the string theory prediction grants the dS universe as much time
as $t_f<{\rm recurrence\,\,time}\sim {\rm e}^{0.5\times 10^{123}}$
yr \cite{Lifetime1}, \cite{Lifetime2}  (the matter of whether it
should be seconds, years or even millenniums is really unessential
for such monstrous numbers). It is possible to lower this value to
the $t_f\sim{\rm e}^{10^{19}}$ yr and even to the limit of
$t_f\sim{\rm e}^{10^{9}}$ yr for models with instantons of Kachru,
Pearson and Verlinde \cite{KPV} and with 2 Klebanov-Strassler (see
\cite{KS}) throats \cite{FLW}. But, nevertheless, even with
assumption that one of those models do describe our Universe, the
magnitude $t_f$ will still be way too large as compared to Page's
$10^{60}$ yr.

To show this, following to Page, suppose that the process of
observation is described by some localized positive operator $A$,
such that application of it to any state $\psi$ leads to positive
central tendency. This implies that every possible observation has
some positive probability of occurrance in the given volume (e.g.,
as a vacuum fluctuation). Therefore, we can treat the observers as
the standart quantum objects. With this in mind, Page has
calculated the action for the brain of a human observer: $S_{{\rm
br}}\sim10^{16}$ $J\times s$, and the probability $p_{{\rm
br}}\sim {\rm e}^{-S/\hbar}\sim {\rm e}^{-10^{50}}$. Then, Page
made an estimation for 4-volume for the brain ($V_4({\rm br})$),
taken in process of making the observation: $V_4({\rm br})\sim{\rm
e}^{331}\,a_{_{\rm Pl}}^4$.

The crucial Page's expression is
\begin{equation}
V_4(t) p_{{\rm ord}}= V_4({\rm br})N, \label{urav}
\end{equation}
where  $V_4(t)$ is the total 4-volume of universe:
\begin{equation}
V_4(t)=\int d^4x\sqrt{-g}\sim \int_{0}^{t} dt a^3(t), \label{V4}
\end{equation}
$ p_{{\rm ord}}$ is the tine of the $V_4(t)$ where all ordered
observations take place and $N$ the total number of such
observations (Following Page we can evaluate $N\sim{\rm e}^{48}$).
If $p_{{\rm ord}}>p_{{\rm br}}$  then we are ordered observers
rather then BBs. Using (\ref{urav}) and (\ref{V4})  we get the
Page's result: $t<10^{60}$ yr if we are not BB.

It were suggested some ways to avoid this conclusions (see
\cite{Linde}, \cite{Vil}, \cite{Buss}). In particularly, Page in
\cite{page-2} has suggested the solution of this problem: Our
vacuum should be rather unstable and should decay within 20 Gyr
(this is possible if the gravitino is superheavy).  He supposed
that the decay of the universe proceeds at the rate, per 4-volume,
of $A$ for the nucleation of a small bubble that then expands at
practically the speed of light, destroying everything within the
causal future of the bubble nucleation event. It is possible if
$A>20 {\rm Gyr}^{-4}$ \cite{page-2}.

In paper we consider the BB problem in frames of phantom cosmology. In Section II the simplest phantom cosmological models with Little Rip and Big singularity are described. The next section is devoted to analysis of compatibility of these models with observational data. The optimal parameters for models are calculated from various observational data such as SNe observations, BAO and Hubble parameter data. In Section IV the Page's mechanism in phantom cosmology is considered. It is showed that Page's solution in this case is incorrect. The mechanism permitting the dominance of BB in these cosmological models is presented in Section V. In Conclusion some outlook is given.

\section{Cosmological models with Little Rip and Big Rip singularity}

Let's try to understand how the BB problem can be solved in frames
of phantom cosmology. We shall see that although Page solution for
BB problem isn't valid in this case but another mechanism of
avoiding of this problem appear.

For simplicity one consider the class of phantom energy models
with equation-of-state (EoS) \be\label{Little}
p=-\rho-\alpha^{2}\rho_{0}\left(\frac{\rho}{\rho_{0}}\right)^{\beta}.
\ee Here $\alpha^{2}$ and $\beta$ are positive constants,
$0\leq\beta\leq1$, $\rho_{0}$ is the phantom energy density in
moment of observation. Therefore for EoS parameter $w_{0}$ we have
simply
$$
w_{0}=-1-\alpha^{2}.
$$
If $\beta=1$ we have simplest phantom model with constant EoS parameter.

\textbf{Phantom model from sub-quantum potential ($\beta=0$).}

The case $\beta=0$ corresponds to model of sub-quantum potential. Sub-quantum potential is interesting idea which allows one to
describe the accelerating of the universe without any dynamics
dark energy. In this model all the speeding-up effects taking
place in our universe are entirely due to the quantum effects
associated with, say, background radiation. The scale factor has
the form
\begin{equation}
a(t)=a_0{\rm e}^{V_{SQ}t^2/4+C_0t}, \label{a}
\end{equation}
where $V_{_{SQ}}=\alpha^{2}\rho_{0}$ is the sub-quantum potential and $C_0$ is some
constant. This model describes the acceleration of the universe
faster the in dS universe but without of the Big Rip singularity.

The time-dependent EoS parameter has the form
\begin{equation}
w(t)=\frac{p}{\rho}=-1-\frac{V_{_{SQ}}}{3\left(V_{_{SQ}}t/2+C_0\right)^2}. \label{w} \end{equation}
Thus $w<-1$ as in phantom universe, but $w\to -1$ as $t\to\infty$.

It s easy to establish the dynamics of universe filled dark energy with EoS (\ref{Little}) using the Friedmann equations. We will examine the future evolution of our universe from the point at which the pressure and density are dominated by the dark energy. One can derive the following link between energy density and time for EoS written in form $p=-\rho-f(\rho)$:
\be
\label{trho}
t = \frac{1}{\sqrt{3}}\int^{\rho}_{\rho_{0}} \frac{d \rho}{\sqrt{\rho} f(\rho)}\, , \quad
x\equiv\sqrt{\rho}\, .
\ee
Thus for $0\leq\beta\leq 1/2$ the singularity doesn't effectively occur: $\rho\rightarrow\infty$ at $t\rightarrow\infty$. For $1/2<\beta\leq 1$ the big rip singularity occurs.

\section{Observational data and models with little rip and big rip singularity}

It is well-known that simplest phantom model describes observational data with sufficient accuracy. We shall see that in principle the dark energy model with EoS (\ref{Little}) for $0<\beta<1$ are compatible with observational data.

We compare the model predictions with data from
from SNe observations, the evolution of the Hubble parameter and baryon acoustic oscillation. The realistic cosmological model should be take
into account that dark energy is not a single component of the
universal energy. We shall see that addition of dark matter allows
one to construct the cosmological models which can be matched with the modern data of
observations.

\textbf{SNe observations.} We use the data for dependence of SNe Ia modulus $\mu$ as function of redshift $z$ from the Supernova Cosmology project \cite{Amman},\cite{Union2}. The theoretical relation for flat universe filled dark energy and matter (for simplicity we neglect the radiation) is
\be
\mu(z)=\mu_{0}+5\lg D_{L}(z).
\ee
where $D_{L}(z)$ is a luminosity distance, that is
\be \label{DLFC}
D_{L}=\frac{c}{H_{0}}(1+z)\int_{0}^{z}
h^{-1}(z)dz,\quad h(z)=\left[\Omega_{m0}(1+z)^{3}+\Omega_{D0}F(z)\right]^{1/2}
\ee
Here, $\Omega_{m0}$ is the total fraction of matter density, $\Omega_{D0}$ the fraction of dark energy energy density, and $H_{0}$ is the current Hubble parameter.
The function $F(z)=\rho_{D}(z)/\rho_{D0}$ can be determined from the continuity equation
\begin{equation}
\dot{\rho}_{D}-3\frac{\dot{a}}{a}(\rho_{D}+p_{D})=0.
\end{equation}

For analysis of observational data one can use the $\chi^2$ statistics. One need to perform a uniform marginalization over free parameter $\mu_{0}$.
Expanding the $\chi^{2}_{SN}$ with respect to $\mu_{0}$ gives
\begin{equation}\label{chi}
\chi^{2}_{SN}=A-2\mu_{0}B+\mu_{0}^{2}C,
\end{equation}
where
$$
A=\sum_{i}\frac{(\mu_{obs}(z_{i})-\mu_{th}(z_{i};\mu_{0}=0))^{2}}{\sigma^{2}_{i}},
$$
$$
B=\sum_{i}\frac{(\mu_{obs}(z_{i})-\mu_{th}(z_{i}))}{\sigma^{2}_{i}},\quad C=\sum_{i}\frac{1}{\sigma^{2}_{i}}.
$$
The expression (\ref{chi}) has a minimum for $\mu_{0}=B/C$ at
$$
\bar{\chi}_{SN}^{2}=A-B^{2}/C.
$$
We will minimize $\bar{\chi}_{SN}^{2}$ instead of ${\chi}_{SN}^{2}$. The 68.3\% confidence level is determined by $\Delta\chi^{2}=\chi^{2}-\chi^{2}_{min}<2.3$ for for two-parametric model. Similarly, the 95.4\% confidence level is determined by $\Delta\chi^{2}=\chi^{2}-\chi^{2}_{min}<6.17$ \cite{Nesseris}.

\textbf{Hubble parameter.} The measurements of $dz/dt$ from fitting stellar population models help to determine the dependence of Hubble parameter as function of redshift
$$
H(z)=-\frac{1}{1+z}\frac{dz}{dt}.
$$
We use the 11 datapoints for H(z) from \cite{Stern} for constraining the model parameters. These data are listed in Table I. The theoretical dependence of the Hubble parameter is
\begin{equation}
H(z)=H_{0}h(z).
\end{equation}

\begin{table}
\label{Table1}
\begin{centering}
\begin{tabular}{|c|c|c|c|}
  \hline
  $z$ & $H_{obs}(z)$ & $\sigma_{H}$  \\
      & km s$^{-1}$ Mpc$^{-1}$        &  km s$^{-1}$ Mpc$^{-1}$     \\
  \hline
  0.090 & 69 & 12 \\
  0.170 & 83 & 8  \\
  0.270 & 77 & 14 \\
  0.400 & 95 & 17 \\
  0.480 & 97 & 62 \\
  0.880 & 90 & 40 \\
  0.900 & 117 & 23 \\
  1.300 & 168 & 17 \\
  1.430 & 177 & 18 \\
  1.530 & 140 & 14 \\
  1.750 & 202 & 40 \\
  \hline
\end{tabular}
\caption{Hubble parameter versus redshift data from \cite{Stern}.}
\end{centering}
\end{table}

The marginalization over the parameter $H_{0}$ can be performed as in a case of analysis of SNe observations. One can minimize the quantity
$$
\bar{\chi}_{H}^{2}=A_{1}-B_{1}^{2}/C_{1},
$$
where
$$
A_{1}=\sum_{i}\frac{H_{obs}(z_{i})^{2}}{\sigma^{2}_{i}},\quad B_{1}=\sum_{i}\frac{H_{obs}(z_{i})}{\sigma^{2}_{i}},\quad C_{1}=\sum_{i}\frac{1}{\sigma^{2}_{i}}.
$$

\textbf{BAO data.} For more precise determination of parameters of cosmological models we use also the BAO data. We use the measurements of the acoustic parameter $A(z)$ from \cite{Blake}, where the theoretically-predicted $A_{th}(z)$ is given by the relation
\begin{equation}\label{Ath}
A_{th}(z)=\frac{D_{V}(z)H_{0}\sqrt{\Omega_{m0}}}{z},
\end{equation}
where $D_{V}(z)$ is a distance parameter defined as
\begin{equation} \label{DV}
D_{V}(z)=\left\{(1+z)^{2}d_{A}^{2}(z)\frac{cz}{H(z)}\right\}^{1/3}.
\end{equation}
Here, $d_{A}(z)$ is the angular diameter distance
\begin{equation} \label{dA}
d_{A}(z)=\frac{y(z)}{H_{0}(1+z)},\quad y(z)=\int_{0}^{z}\frac{dz}{h(z)}.
\end{equation}

Using Eqs.~(\ref{Ath})-(\ref{dA}) we have
\begin{equation}
A_{th}(z)=\sqrt{\Omega_{m0}}\left(\frac{y^{2}(z)}{z^{2}h(z)}\right).
\end{equation}
Using the WiggleZ $A_{obs}(z)$ data from Table 3 of \cite{Blake}, we compute $\chi^{2}_{A}$ as
\begin{equation}
\chi^{2}_{A}=\Delta \textbf{A}^{T}(C_{A})^{-1}\Delta\textbf{A}.
\end{equation}
Here, $\Delta\textbf{A}$ is a vector consisting of differences, $\Delta A_{i}=A_{th}(z_{i})-A_{obs}(z_{i})$ and $C^{-1}_{A}$ is the
inverse of the $3 \times 3$ covariance matrix given in Table 3 of \cite{Blake}.

\textbf{Optimal parameters for little rip model with $\beta=0$.}

For sub-quantum potential
\be
F(z)=1+3(1+w_{0})\ln(1+z)
\ee
Here
$$
w_{0}=-1-\frac{V_{SQ}}{3C_{0}^{2}}.
$$

For $w_{0}=-1.024$ the parameter $\chi^{2}=\bar{\chi}^2_{SN}+\bar{\chi}^2_{H}+\chi^{2}_{A}$ reaches the minimal value $560.92$ ($\Omega_{D}=0.713$). In the case of $\Lambda$CDM cosmology we have $\chi^{2}_{min}=561.31$ ($\Omega_\Lambda=0.712$). The value of $w_{0}$ lies in interval $-1.125\leq w_{0}\leq-1.000$ with 68 \% confidence level and in interval $-1.185\leq w_{0}\leq-1.000$ with 95\% confidence level.

For $\beta>0$ the same picture take place. The parameter
$\alpha^2<\approx 0.2$ with 95\% confidence level for $0<\beta<1$
(for our estimations such accuracy for bound
$\alpha^2_{\mbox{max}}$ is sufficient).

\section{Page's solution of BB problem and phantom cosmology}

Let's show that in universe filled phantom energy with EoS (\ref{Little}) Page's solution of BB problem is incorrect. One consider the case of sub-quantum potential. As we shall see the consideration of this simple case is sufficient for general result.

Let suppose that the present value of
$w(0)=w_0=-1-\epsilon/3$ with $\epsilon=V_{_{SQ}}/H_0^2$ and
the present value of Hubble roots is $H_0$. Then one can use
(\ref{a}) and (\ref{w}) to present the scale factor in the form
\begin{equation}
a(t)=a_0\exp\left(\frac{H_0t}{4}\left(\epsilon
H_0t+4\right)\right). \label{a1}
\end{equation}
Substituting (\ref{a1}) in (\ref{V4}) one get
\begin{equation}
V_4(t)=\frac{i ca_0^3{\rm
e}^{-3/\epsilon}}{H_0}\sqrt{\frac{\pi}{3\epsilon}}\left({\rm
erf}\left[i\left(\sqrt{\frac{3}{\epsilon}}+\frac{\sqrt{3\epsilon}
H_0t}{2}\right)\right]-{\rm{erf}}\left[i\sqrt{\frac{3}{\epsilon}}\right]\right).
\label{VV4}
\end{equation}
For large value of $H_0t=\tau$ one can estimate the expression
(\ref{VV4}) as
\begin{equation}
V_4(\tau)\sim \frac{2}{3H_{0}\epsilon \tau}\exp\left[\frac{3\epsilon\tau^2}{4}\right].
\label{V-crude}
\end{equation}
Now, let consider the Page's bubble-killer. Let us take the case
in which the decay of the universe proceeds by the nucleation of a
small bubble that then expands at practically the speed of light,
destroying everything within the causal future of the bubble
nucleation event. Suppose that the bubble nucleation rate, per
4-volume, is $A$. The probability that the spacetime would have
survived to the event $Q$ is
$$
P(Q)={\rm e}^{-AV_4(Q)},
$$
where $V_4(Q)$ is the spacetime 4-volume to the past of the event
$Q$ in the background spacetime.

The requirement that there not be an infinite expectation value of
vacuum fluctuation observations within a finite comoving 3-volume
 is the requirement that
\begin{equation}
\int d^4x \sqrt{-g}P(Q)<\infty. \label{req}
\end{equation}
In the case of the dS universe (as Page shown) the requirement
(\ref{req}) is valid if and only if $A>20 {\rm Gyr}^{-4}$
\cite{page-2}. Now let consider the universe with sub-quantum
potential and scale factor (\ref{a1}). One can use the the
conformal time $\eta$
$$
\eta=\frac{cT}{a_0}\sqrt{\frac{\pi}{\epsilon}}{\rm
e}^{1/\epsilon}\left[{\rm
erf}\left(\frac{\epsilon\tau+2}{2\sqrt{\epsilon}}\right)-{\rm
erf}\left(\frac{1}{\sqrt{\epsilon}}\right)\right],
$$
where $T=1/H_0$. Then we have the flat FRW metric
$$
ds^2=a^2(\eta)\left(d\eta^2-dr^2-r^2d\Omega^2\right),
$$
and
$$
V_4(Q)=\frac{4\pi}{3}\int_0^{\eta} d\eta'
a^4(\eta')\left(\eta-\eta'\right)^3.
$$
Thus the probability that the spacetime would have survived to the
event $Q$ is
\begin{equation}
P(Q)=\exp\left[-\frac{A}{A_m}\int_0^{\tau} dx {\rm e}^{3x(\epsilon
x+4)/4}\left({\rm
erf}\left(\frac{\epsilon\tau+2}{2\sqrt{\epsilon}}\right)- {\rm
erf}\left(\frac{\epsilon x+2}{2\sqrt{\epsilon}}\right)
\right)\right],
 \label{prob}
\end{equation}
where
$$
A_m=\frac{3\epsilon^{3/2}{\rm e}^{-3/\epsilon}}{4\pi^{5/2}(cT)^4}.
$$
Using the expression (\ref{prob}) we conclude that the requirement
(\ref{req}) is valid if and only if
\begin{equation}
J\equiv \int_{1/\sqrt{\epsilon}}^{+\infty}
d\xi\exp\left[3\xi^2-\frac{A'}{A_m}\int_{1/\sqrt{\epsilon}}^{\xi}
dz {\rm e}^{3z^2}\left({\rm erf}\xi-{\rm erf}
z\right)^3\right]<\infty, \label{req-1}
\end{equation}
with $A'=2A{\rm e}^{-3/\epsilon}/\sqrt{\epsilon}$.

One can show that the the requirement (\ref{req-1}) can't be
valid, since the second integral in (\ref{req-1}) is finite as
$\xi\to\infty$. Thus $J=\infty$ for any values of  the bubble
nucleation rate, per 4-volume  (i.e. $A$). It is possible to
describe a situation so: the universe with sub-quantum potential
extends so fast that even the bubble growing with speed of light
can't destroy this one. Of course this conclusion is right for
dark energy model with EoS (\ref{Little}) for arbitrary
$0<\beta\leq 1$: the expansion of universe at $\beta>0$ occurs
more quickly than in a case of sub-quantum potential.

\section{The possibility of resolution of BB problem in phantom cosmology}

Thus, the Page's mechanism does not work. Nevertheless, there is
other way to decide the paradox of BB. Page's doomsday
argument is true only in an universe which contains
human-observers. But such observers can exist in the universe
filled with phantom energy only up to a certain period. The upper
bound $t_{\rm{max}}$ can be obtained by the Bekenstein bound (see
below). Therefore if $V_4(t_{\rm{max}})>V_4({\rm cr})$ then we
have problems with such models in the light of Page's doomsday
argument. As we shall see, for $\beta>\beta_{min}$  this is the case only if the
``fine-tuning'' of $w$ take place. In the case of general position
(i.e. without ``fine-tuning'') cosmological models with phantom
energy don't suffer from the Page's doomsday argument therefore
such models are more realistic then models with $\Lambda$-term.

The Bekenstein bound \cite{11} shows that the total amount of
information, which can be stored in region of radius $R$ is
$I<I_{m}=2\pi R M c/(\hbar \ln 2)$ thus
\begin{equation}
I<I_{m}=2.59\times 10^{38}\,\left(\frac{M}{1\,\,{\rm
g}}\right)\left(\frac{R}{1\,\,{\rm cm}}\right)\,\,{\rm
bits}. \label{BB}
\end{equation}
Substituting $\rho_0=\rho_c=10^{-29}$ ${\rm g}/{\rm cm}^3$,
$R=c/H_0$, and $H_0\sim 70$ km/s/Mps (the current measured value
of the Hubble constant) results in
\begin{equation}
 I_m=2.97\times 10^{122}\,{\rm bits}
 \label{naiv}.
\end{equation}

In general case, the horizon distance can be calculated as
\begin{equation}
R_c(t)=a(t)\int_t^{t_{_f}}\frac{c dt'}{a(t')}. \label{Rc}
\end{equation}

\textbf{Model with sub-quantum potential.} In the case of sub-quantum model we get
\begin{equation}
R_c(\tau)=R_{dS}{\rm
e}^{1/\epsilon}\sqrt{\frac{\pi}{\epsilon}}{\rm
e}^{\tau(\epsilon\tau+4)/4}\left[1-{\rm
erf}\left(\frac{\epsilon\tau+2}{2\sqrt{\epsilon}}\right)\right],
\label{Rc-1}
\end{equation}
where $\tau=H_0t$, $R_{dS}=c/H_0$. At present time
$$
R_c(0)=R_{dS}{\rm
e}^{1/\epsilon}\sqrt{\frac{\pi}{\epsilon}}\left(1-{\rm
erf}\left(\frac{1}{\sqrt{\epsilon}}\right)\right).
$$
For large values of $\tau$ we get more simple expression
\begin{equation}
R_c(t)\sim\frac{2R_{dS}}{\epsilon \tau}, \label{Rc-simple}
\end{equation}
so $R_c(t)\to 0$ as $t\to\infty$.

Now, using \ref{Rc-simple}, (\ref{V-crude}) and the Page
requirement
$$
V_4(t)<NV_4({\rm br}){\rm e}^{S_{\rm br}/\hbar},
$$
we get
\begin{equation}
\frac{3R_{dS}^{2}}{\epsilon R_c^2}<\frac{S_{\rm br}}{\hbar}.
\label{req-2}
\end{equation}
that result in low limit on the parameter $\epsilon$:
\begin{equation}
\epsilon>\epsilon_{\rm min}=\frac{3R_{dS}^{2}\hbar}{R_{min}^{2}S_{\rm
br}}. \label{ep-min}
\end{equation}
From the (\ref{BB}) one can obtain the low limit on the value of
$R_c$:
\begin{equation}
R_c\ge R_{\rm min}\frac{\hbar I \ln 2}{2\pi Mc}, \label{limit}
\end{equation}
where $I$ is the amount of information encoded in a
human-observer.

Using (\ref{ep-min}) and (\ref{limit}) we get
\begin{equation}
\epsilon_{\rm min}\le 3.33\times 10^{83}\frac{M^2}{I^2}. \label{final}
\end{equation}
For the $M=100$ kg and $I\sim 10^{45}$ bits (the upper amount of
information encoded in a human-observer) we get
the (\ref{final})
\begin{equation}
\epsilon>\epsilon_{\rm min}\approx 3300, \label{ho-ho}
\end{equation}
or
\begin{equation}
w_0<-1100 \label{ho-ho-ho}
\end{equation}
From previous section one can see that this bound on $w_{0}$
conflicts with observational data. Therefore one can conclude that
in universe with sub-quantum potential the BB dominate in
comparison with ordinary observers.

\textbf{Simplest phantom model.}However it is interesting to note
that the same mechanism permitting to avoid dominance of BB in
simplest phantom model with constant EoS parameter. Seemingly the
big rip results in the problems which are even greater then in the
de Sitter case. Really, $a(t)\to+\infty$ since $t\to t_{_f}$ so
$V_4\to+\infty$ and therefore $V_4$ would necessarily overrun the
$V_4({\rm cr})={\rm e}^{10^{50}} a_{{\rm Pl}}^4$ and it would make
it more quickly then in the de Sitter case.

Integration of the Einstein-Friedmann equation for the flat universe filled phantom energy with $w_{0}=-1-\alpha^{2}$ results
in
\begin{equation}
\begin{array}{cc}
\displaystyle{
a(t)=\frac{a_0}{\left(1-\xi t\right)^{2/3\alpha^{2}}},}\\
\displaystyle{
\rho(t)=\rho_0\left(\frac{a(t)}{a_0}\right)^{3\alpha^{2}}=\frac{\rho_0}{(1-\xi
t)^2}}, \label{1}
\end{array}
\end{equation}
where $\xi=\alpha^{2}\sqrt{6\pi G\rho_0}$. We choose $t=0$ as the
present time, $a_0\sim 10^{28}$ cm and $\rho_0$ to be the present
values of the scale factor and the density. There, if
$t=t_{_f}=1/\xi$, we automatically get the big rip. Using
(\ref{1}) one can calculate $V_4(t)$, as
\begin{equation}
\displaystyle{ V_4(t)=\frac{\alpha^{2} c
a_0^3}{\xi(2-\alpha^{2})}\left(\frac{1}{\left(1-\xi
t\right)^{(2-\alpha^{2})/\alpha^{2}}}-1\right)}. \label{V4-1}
\end{equation}

Substituting (\ref{1}) into the (\ref{Rc}) gives
\begin{equation}
R_c(t)=\frac{3c\alpha^{2}(1-\xi t)}{(2+3\alpha^{2})\xi}, \label{Rcph}
\end{equation}
for the case of phantom field. This, of course, means that
$R_c(t)\to 0$ as $t\to t_{_f}$. Finally, the Bekenstein Bound
(\ref{BB}) results in
\begin{equation}
I<I_m({ t})\sim \frac{2.74 R^4_c({ t})\rho_0}{\left(1-\xi{
t}\right)^2}\times 10^{43}. \label{Imm}
\end{equation}
Using the superior limit of amount of information encoded in a
human-observer one get
(using (\ref{Imm}) and (\ref{Rcph}))
\begin{equation}
\displaystyle{ \eta\equiv 1-\xi
t>\frac{0.67\xi^2(2+3\alpha^{2})^2}{c^2\alpha^2\sqrt{\rho_0}}.}
\label{ner-1}
\end{equation}
On the other hand, using (\ref{V4}) and the condition
$V_4(t)<V_4({\rm cr})$ one get
\begin{equation}
\displaystyle{ \eta<\left(\frac{\alpha^{2} a_0^3
c}{\xi(2-\alpha^{2})V_4({\rm cr})+\alpha^{2}
a_0^3c}\right)^{\alpha/(2-\alpha)}}. \label{ner-2}
\end{equation}
Combining (\ref{ner-1}) and (\ref{ner-2}) we have
\begin{equation}
\displaystyle{ \frac{4\pi G(2+3\alpha^{2})^2\sqrt{\rho_0}}{c^2}<\left(\frac{a_0^3
c}{(2-\alpha^{2})V_4({\rm cr})\sqrt{6\pi G\rho_0}+
a_0^3c}\right)^{\alpha^{2}/(2-\alpha^{2})}}. \label{nerv}
\end{equation}
Choosing $\alpha^{2}\ll 1$ one get
$$
-88.48<\log\eta<-\frac{\alpha^{2}}{2}\times 10^{50},
$$
so
\begin{equation}
\alpha^{2}<1.77\times 10^{-48}. \label{ep}
\end{equation}
It is very unlikely that $\alpha^{2}$ is so small accurate to
$10^{-48}$! Therefore it is unlikely that we have some problems
with Page's doomsday argument in phantom universe. It is
interesting that if (\ref{ep}) is the case then $t_{_f}>5\times
10^{57}$ yr. If $\beta>1$ our conclusion remains unchanged.

Let consider this situation by another way. The total  history of
universe can be divided on two parts. The first one is the
``observable universe'' which can contain human-observers and the
second ``unobservable universe'' where human-observers can't
exist. The $t_{{\rm observ}}$ can expressed from the (\ref{ner-1})
while the total lifetime of universe $t_{{\rm
total}}=t_{_f}=1/\xi$. Therefore
$$
\frac{t_{{\rm observ}}}{t_{{\rm total}}}<1-0.93\times
10^{-39}\left(2+3\alpha\right)^2.
$$
It is easy to see that for $0<\alpha^{2}<1$ this ratio will be 1
accurate to $10^{-39}$, so the universe is ``observable'' one
during virtually whole it's history. On the other hand, as we
seen above, in ``observable universe''   filled with phantom
energy our ordered observations would be highly typical in
contrast to universe filled with positive cosmological constant
where our ordered observations would be highly atypical.

\textbf{Model (\ref{Little}) with $0<\beta<1/2$.} Now let's
consider the case of intermediate cosmological dynamics between
model of sub-quantum potential and big rip ($\beta>1/2$). In this
case the scale factor has a form
\begin{equation}
a(t)=a_{0}\exp\left(3^{-1}\alpha^{-2}(1-\beta)^{-1}(3^{1/2}\alpha^{2}(1/2-\beta)\tau+1)^{\gamma}-1\right), \quad \gamma=\frac{1-\beta}{1/2-\beta}.
\end{equation}
For $\tau>>3^{-1/2}\alpha^{-2}(1/2-\beta)^{-1}$ one obtain simply
\begin{equation}
a(t)\approx\exp\left(B(\beta)\alpha^{2(\gamma-1)}\tau^{\gamma}\right),\quad
B(\beta)=\frac{3^{\gamma/2-1}}{1-\beta}\left(1/2-\beta\right)^{\gamma}.
\end{equation}

For horizon distance one obtain
\begin{equation}
R_{c}(\tau)\approx\frac{R_{dS}}{\gamma B(\beta)\alpha^{2(\gamma-1)}\tau^{\gamma-1}}.
\end{equation}
and for $\ln V_{4}(\tau)$
\begin{equation}
\ln V_{4}(\tau)\approx 3B(\beta)\alpha^{2(\gamma-1)}\tau^{\gamma}\approx \frac{3}{\alpha^{2}}\left(\frac{R_{dS}}{R_{c}}\right)^{\frac{\gamma}{\gamma-1}}\left(B(\beta)\gamma^\gamma\right)^{\frac{1}{1-\gamma}}.
\end{equation}
From Page requirement one obtain as in a case of sub-quantum potential the low limit on $\alpha^{2}$:
\begin{equation}\label{LIMIT}
\alpha^2>3\frac{\hbar}{S_{br}}\left(\frac{R_{dS}}{R_{\rm
min}}\right)^{\frac{\gamma}{\gamma-1}}\left(B(\beta)\gamma^\gamma\right)^{\frac{1}{1-\gamma}}.
\end{equation}

Taking into account the maximal limit on $\alpha^2$ from
observations ($\alpha^2<\approx 0.2$)  one can conclude that
dominance of BB occurs in models with $0\leq\beta\leq 0.08$. The
limit on $\alpha^2$ from (\ref{LIMIT}) decreases rapidly with
growing $\beta$. For $\beta=0.161$ one has $\alpha^2_{\rm
min}=10^{-5}$, for $\beta=0.255$ - $\alpha^2_{\rm min}=10^{-10}$.
Therefore for $\beta>\beta_{min}$ BB dominate only if we have the
``fine-tuning'' of EoS parameter.

Therefore one can see that model with EOS (\ref{Little}) is free
from BB for sufficiently wide interval of $\beta$.

\section{Conclusion}

The problem of Boltzmann brains in frames of phantom models with
big rip and little rip is considered. Among these models there are
models satisfying to observational tests and free from BB problem.
The resolution of BB problem is achieved due to Bekenstein bound
leading to separation of universe history on two parts in fact.
The fraction of BB in ``observational'' part (in which the human
observers can exist) is negligible in comparison with ordered
observers. The analysis of observational data shows that allowable
range of parameters includes such values at which the ``observable
universe'' consists of ordered observations mainly.

\acknowledgments

This work is supported by project 14-02-31100 (RFBR,
Russia) (AVA).


\begin{thebibliography}{99}

\bibitem{Riess} A.G. Riess et al., Astron. J. {\bf 116}, 1009 (1998).

\bibitem{Perlmutter} S. Perlmutter et al., Ap. J. {\bf 517}, 565 (1999).

\bibitem{Dark-1} E. Copeland, M. Sami and S. Tsujikawa, Int. J. Mod. Phys. D {\bf 15}, 1753 (2006).

\bibitem{Dark-2} R. Caldwell and M. Kamionkowski, Ann. Rev. Nucl. Part. Sci. {\bf 59}, 397 (2009).

\bibitem{Dark-3} R. Durrer and R. Maartens, Gen. Rel. Grav. {\bf 40}, 301 (2008).

\bibitem{Dark-4} J. Frieman and M. Turner, Ann. Rev. Astron. Astrophys. {\bf 46}, 385 (2008).

\bibitem{Dark-5} A. Silvestri and M. Trodden, Rept. Prog. Phys. {\bf 72}, 096901 (2009).

\bibitem{Dark-6} M. Li, X. Li, S. Wang and Y. Wang, Commun. Theor. Phys. {\bf 56}, 525 (2011).

\bibitem{Kowalski} M. Kowalski, Ap. J. {\bf 686}, 74 (2008).

\bibitem{PDP} Review of Particle Physics, 102 (2010).

\bibitem{Amman} R. Amanullah et al., Ap. J. {\bf 716}, 712 (2010).

\bibitem{Carrol} S.M. Carroll, M. Hofman and M. Trodden, Phys. Rev. D {\bf 68}, 023509 (2003).

\bibitem{Starobinsky} A.A.~Starobinsky, Grav. Cosmol. {\bf 6}, 157 (2000).

\bibitem{Caldwell} R.R.~Caldwell, Phys.\ Lett. B {\bf 545} 23 (2002).

\bibitem{Frampton} P.H. Frampton and T. Takahashi, Phys. Lett. B {\bf 557}, 135 (2003).

\bibitem{Diaz} P.F. Gonz$\acute{a}$lez-Di$\acute{a}$z, Phys. Lett. B {\bf 586}, 1 (2004).

\bibitem{Nojiri} S. Nojiri and S.D. Odintsov, Phys. Rev. D {\bf D 70}, 103522 (2004).

\bibitem{Nojiri-3} S. Nojiri, S.D.~Odintsov, and S. Tsujikawa, Phys. Rev. D {\bf 71}, 063004 (2005).

\bibitem{Stefanic} H. Stefancic, Phys. Rev. D {\bf 71}, 084024 (2005).

\bibitem{Frampton-2} P.H. Frampton, K.J. Ludwick and R.J. Scherrer, arXiv:1106.4996.

\bibitem{Frampton-3} P.H. Frampton, K.J. Ludwick and R.J. Scherrer, arXiv:1112.2964.

\bibitem{Wei} H. Wei and R.G. Cai, Phys. Rev. D {\bf 72}, 123507 (2005).

\bibitem{Chimento} L.P. Chimento, Phys. Rev. D {\bf 81}, 043525 (2010).

\bibitem{Cai} R.G. Cai and Q.P. Su, Phys. Rev. D {\bf 81}, 103514 (2010).

\bibitem{ito} Y. Ito, S. Nojiri and S.D. Odintsov, arXiv:1111.5389.

\bibitem{Frampton-4} P.H. Frampton, K.J. Ludwick, S. Nojiri, S.D. Odintsov and R.J. Scherrer, arXiv:1108.0067.

\bibitem{page-1} D.N. Page, J. Corean Phys. Soc. \textbf{49}, 711 (2006).

\bibitem{Lifetime1} N. Goheer, M. Kleban, and L. Susskind, JHEP \textbf{07}, 056 (2003).

\bibitem{Lifetime2} S. Kachru, R. Kallosh, A. Linde, S.P. Trivedi, Phys. Rev. D \textbf{68}, 046005 (2003).

\bibitem{KPV} S. Kachru, J. Pearson, and H. Verlinde, JHEP \textbf{06}, 021 (2002).

\bibitem{KS} I.R. Klebanov and M.J. Strassler, JHEP \textbf{08}, 052 (2000).

\bibitem{FLW} A.R. Frey, M. Lippert, and B. Williams, Phys. Rev. D \textbf{68}, 046008 (2003).

\bibitem{Linde} A. Linde, hep-th/0611043.

\bibitem{Vil} A. Vilenkin, hep-th/0611271.

\bibitem{Buss} R. Bousso and B. Freivogel, hep-th/0610132.

\bibitem{page-2} D.N. Page, hep-th/0610079; D.N. Page, hep-th/0612137.

\bibitem{Union2} http://supernova.lbl.gov

\bibitem{Nesseris} S. Nesseris and L. Perivolaropoulos, Phys. Rev. D {\bf 72}, 123519 (2005).

\bibitem{Stern} D. Stern et al., JCAP {\bf 1002}, 008 (2010).

\bibitem{Blake} D. Blake et al., Mon. Not. Roy. Astr. Soc. {\bf 418}, 1707 (2011).

\bibitem{11}  J.D. Bekenstein, Phys. Rev. Lett. \textbf{46},  623 (1981).

\end{thebibliography}
\end{document}